# Investigation of alpha particle induced reactions on natural silver in the 40-50 MeV energy range

**F. Ditrói[1*], S. Takács[1], H. Haba[2], Y. Komori[2], M. Aikawa[3], M. Saito[4], T. Murata[5]**

[1]Institute for Nuclear Research, Hungarian Academy of Sciences, Debrecen, Hungary
[2]Nishina Center for Accelerator-Based Science, RIKEN, Wako, Japan
[3] Faculty of Science, Hokkaido University, Sapporo, Japan
[4]Graduate School of Biomediacal Science and Engineering, Hokkaido University, Sapporo, Japan
[5]School of Science, Hokkaido University, Sapporo, Japan

## Abstract

Natural silver targets have been irradiated by using a 50 MeV alpha-particle beam in order to measure the activation cross sections of radioisotopes in the 40-50 MeV energy range. Among the radio-products there are medically important isotopes such as $^{110m}$In and $^{111}$In. For optimizing the production of these radioisotopes and regarding their purity and specific activity the cross section data for every produced radioisotope are important. New data are measured in this energy range and the results of some previous measurements have been confirmed. Physical yield curves have been calculated by using the new cross section data completed with the results from the literature.

* Corresponding author: ditroi@atomki.hu

## 1. Introduction

As silver is a basic material for producing radioisotopes for medical and for industrial applications it is of fundamental importance to know the corresponding nuclear data of the produced radioisotopes (cross section and yield) with reasonable accuracy. Among the produced radioisotopes e.g. $^{111}$In plays an important role in nuclear medicine as diagnostic isotope [1], but its relatively long half-life allows also industrial and research applications [2]. $^{109}$Cd is an interesting radioisotope from the point of view of medical applications as a parent of $^{109m}$Ag [3-5]. Nuclear data for alpha particle induced reactions on silver have already been measured by several authors [6-12]. In this work in the frame of a higher energy alpha experiment series by using stacked foil technique we covered the energy range 40-50 MeV using high purity silver foils as targets in order to complement the failing data in this energy range and resolve the contradictions between the existing data.

## 2. Experimental

The irradiations have been performed on a dedicated beam line of the K70-MeV AVF cyclotron of the RIKEN RI Beam Factory by using an $E_\alpha$ = 50.73±0.3 MeV beam. The exact beam energy was determined by using the time of flight (TOF) setup [13]. The well-established stacked foil technique was used for assessing the excitation function of the different nuclear reactions. A combined stack was constructed, it contained foils for two experiments, i.e. Ag+$\alpha$ and Ni+$\alpha$ cross section measurements. In this work we discuss only the results on silver. The used target foils were high purity (at least 99.99%) Goodfellow © foils with the following thicknesses: Ag: 30 and 8.25 μm. The actual thickness of the foils, which was in general different from the nominal one, was determined by weighting the whole metal sheet purchased, and measuring the exact lateral size, from which an average thickness was determined, assuming that the whole sheet was homogeneous and had even thickness all over the sheet. The $^{nat}$Ti($\alpha$,x)$^{51}$Cr reaction on titanium foils (Ti: 10.9 μm) was used as a monitor reaction to check and correct the beam intensity and energy degradation through the whole stack. The foils are ordered in groups in such a way that we could compensate or avoid the activity loss or excess activity because of recoil effect of the radioisotope in question. Only those foils were involved in the final evaluation, where the recoil from the particular target foil was compensated by the recoil from the preceding foil (same material), or it was measured together with the following foil (different material) if the preceding foil was not the same material. The Ag foils were arranged in one block, i.e. one after each other, because for silver we were interested in the high energy part. (50 - 40 MeV range). The first 10 foils of the stack were silver. The first Ag foil was thick (30 μm), because in this case the recoil was proportionally small, all the others were thinner (8.25 μm). In the case of the thin Ag foils the recoil effect was compensated, because foil from the same material, Ag was before each target foil. The irradiation lasted for one hour at 200 nA beam current. After a short cooling time the stack was disassembled into single foils or pairs of foils and the gamma-ray measurements begun. A HPGe semiconductor detector based spectrometer was used for the gamma-ray measurements. Three series of measurements have been performed on silver targets with different cooling times from 20 hours to 10 days. The spectra were evaluated later by using the automatic software [14] and in special cases (overlapping multiple and/or weak peaks) manual evaluation by using a home-developed evaluation software [15] was also performed. The activity of the Ti monitor foils, which were also inserted into the stack in pairs, in order to compensate the recoil effect of the lower energy foil of the pair, were also measured later in order to fit the beam energy and intensity. The whole excitation function of the $^{nat}Ti(\alpha,x)^{51}Cr$ monitor reaction was re-measured and the results were compared with the recommended values of the IAEA monitor reaction database [16]. The final beam energy and intensity were adjusted according to the best fit with the recommended values [17]. Based on these results the initial parameters for all calculations were set. The used nuclear data are listed in Table 1 for all measured radioisotopes.

The uncertainties of the single cross section values were estimated by calculating square root of the sum in quadrature of all single contributions [18]: beam current (5%), target thicknesses (3%), detector efficiency (5%), nuclear data (3%), peak area and counting statistics (1-20%), the overall uncertainty in the results (cross sections) was 7-20 %.

**Table 1** Nuclear data for the radioisotopes produced [19, 20]

| Isotope<br>spin<br>level energy(keV) | Half-life | Decay mode | $E_\gamma$<br>(keV) | $I_\gamma$<br>(%) | Contributing reactions | Q-value<br>(MeV) |
|---|---|---|---|---|---|---|
| $^{111g}$In<br>9/2$^+$ | 2.8047 d | ε: 100% | 171.28<br>245.35 | 90.7<br>94.7 | $^{109}$Ag(α,2n) | -14.05 |
| $^{110g}$In<br>7$^+$ | 4.9 h | ε: 100% | 657.75<br>884.68<br>937.48 | 98<br>93<br>68.4 | $^{107}$Ag(α,n)<br>$^{109}$Ag(α,3n) | -7.58<br>-24.04 |
| $^{110m}$In<br>2$^+$<br>62.084 | 69.1 min | ε: 100%<br>β$^+$: 61.3% | 657.75 | 97.74 | $^{107}$Ag(α,n)<br>$^{109}$Ag(α,3n) | -7.58<br>-24.04 |
| $^{109g}$In<br>9/2$^+$ | 4.159 h | ε: 100%<br>β$^+$: 4.64% | 203.3 | 74.2 | $^{107}$Ag(α,2n)<br>$^{109}$Ag(α,4n) | -15.63<br>-32.09 |
| $^{108g}$In<br>7$^+$ | 58 min | ε: 100%<br>β$^+$: 24.8% | 632.9<br>875.4 | 100<br>100 | $^{107}$Ag(α,3n)<br>$^{109}$Ag(α,5n) | -26.75<br>-44.09 |
| $^{111g}$Ag<br>½$^-$ | 7.45 d | β$^-$: 100% | 245.4 | 1.24 | $^{109}$Ag(α,2p) | -12.66 |
| $^{110m}$Ag<br>6$^+$<br>117.595 | 249.76 d | IT: 1.33%<br>β$^-$: 100.4% | 657.76<br>763.94<br>884.68<br>937.49 | 95.61<br>22.6<br>75<br>35 | $^{109}$Ag(α,2pn) | -21.49 |
| $^{106m}$Ag<br>6$^+$<br>89.667 | 8.28 d | ε: 100% | 450.98<br>717.34<br>1045.83<br>1527.65 | 28.2<br>28.9<br>29.6<br>16.3 | $^{107}$Ag(α,2p3n)<br>$^{109}$Ag(α,2p5n) | -37.83<br>-54.29 |
| $^{105g}$Ag<br>1/2$^-$ | 41.29 d | ε: 100% | 280.44<br>344.52<br>443.37 | 30.2<br>41.4<br>10.5 | $^{107}$Ag(α,2p4n)<br>$^{109}$Ag(α,2p6n) | -45.77<br>-62.23 |
| $^{109}$Cd<br>5/2$^+$ | 461.9 d | ε: 100% | 88.03 | 3.64 | $^{107}$Ag(α,2p5n)<br>$^{109}$Ag(α,2p7n) | -55.80<br>-72.26 |

Increase the Q-values if compound particles are emitted by: np-d, +2.2 MeV; 2np-t, +8.48 MeV; n2p-$^3$He, +7.72 MeV; 2n2p-α, +28.30 MeV.
Decrease Q-values for isomeric states with level energy of the isomer
Abundances: $^{107}$Ag (51.83%), $^{109}$Ag (48.17%)

## 3. Theoretical model calculations

We wanted to test the prediction capability of the theoretical nuclear reaction model codes. Calculations for the measured cross sections were made by using the modified TALYS 1.8 code [21] presented in the TENDL-2017 on-line library [22]. Production cross sections with the latest version of the EMPIRE code [23], (EMPIRE 3.2 (Malta) [24]), which contained the latest reference input parameters library RIPL-3 [25], were also given for comparison. The code run with the default input parameters by considering all possible reactions involved, including emission of complex particles above the reaction thresholds at the given bombarding energies.

## 4. Results and discussion

The experimental cross sections deduced for the $^{nat}Ti(\alpha,x)^{51}Cr$ monitor reaction were compared with the recommended cross section from the IAEA monitor reaction database [16]. After very small adjustment of the beam intensity and using the actual foil thicknesses, an excellent agreement was found between the recommended and the measured data (Fig. 1).

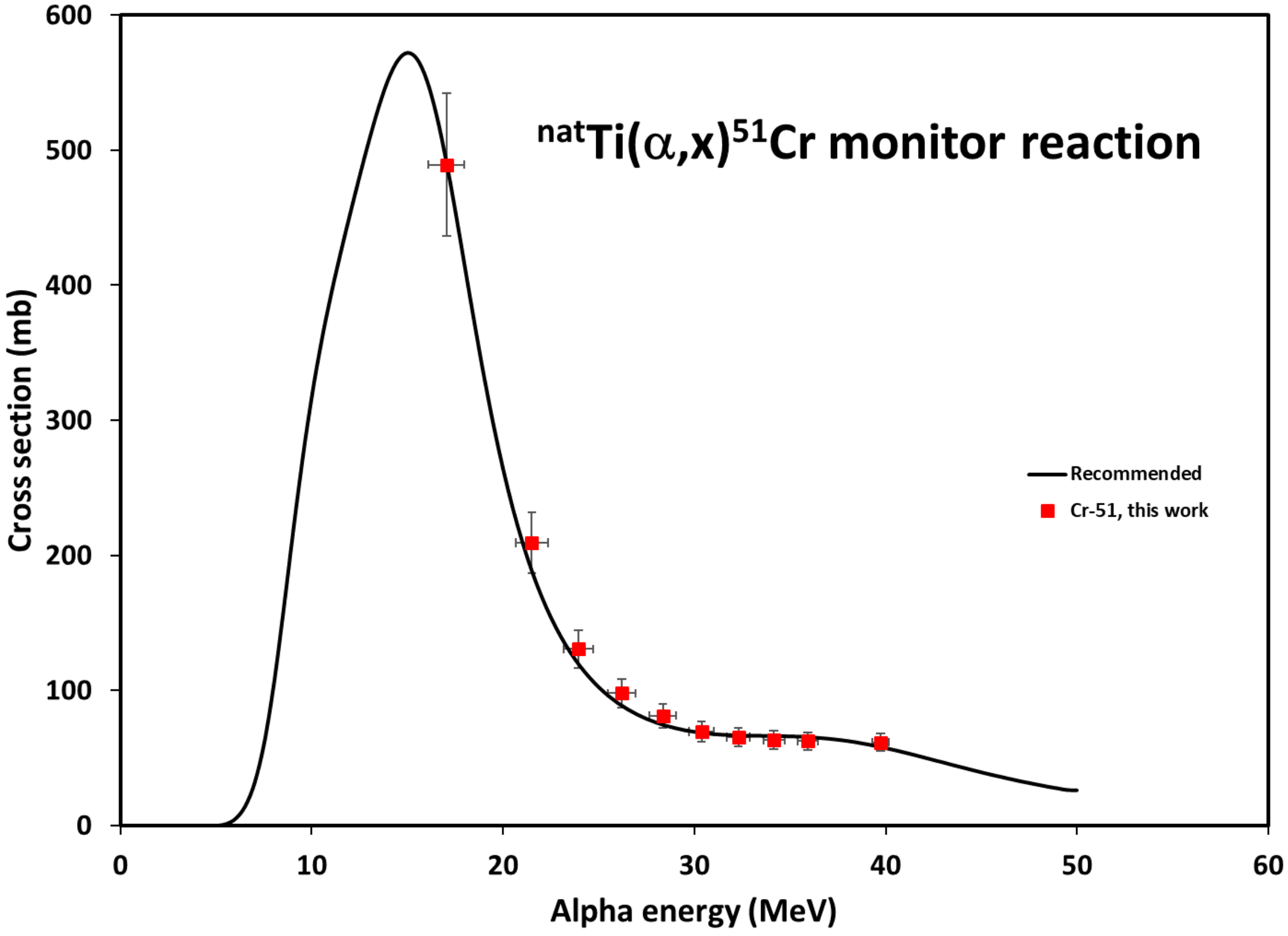

Fig. 1 Re-measured cross section data of the $^{nat}Ti(\alpha,x)^{51}Cr$ monitor reaction compared with the recommended data

From Fig. 1 it is seen that the first monitor foil in the stack is placed behind the last silver target foil. Since we have a good agreement among the measured and the recommended values for the monitor reaction in the low energy segment of the stack, the energy degradation calculation within the high energy segment of the stack, the silver foils of the stack, was correct. The new results for all measured radioisotopes are presented in Figs. 2-11 and in Tables 2-3.

### 4.1 $^{nat}Ag(\alpha,x)^{111}In$ nuclear reaction

This reaction is possible only on the $^{109}Ag$ target isotope by emission of two neutrons. The $^{107}Ag$ target isotope does not play a role in this case. The ground state has a long enough half-life to measure the activity of the foils in all the three spectrum series. The short-lived (7.6 min) meta-stable state decays completely to the ground state by internal conversion. The results given in Fig 2. are deduced from the intense 171.28 keV gamma-line, which is free from interferences. All the three series gave consistent results. The other intense, 245.35 keV gamma-line was excluded from the evaluation because of interferences with other produced radioisotopes.

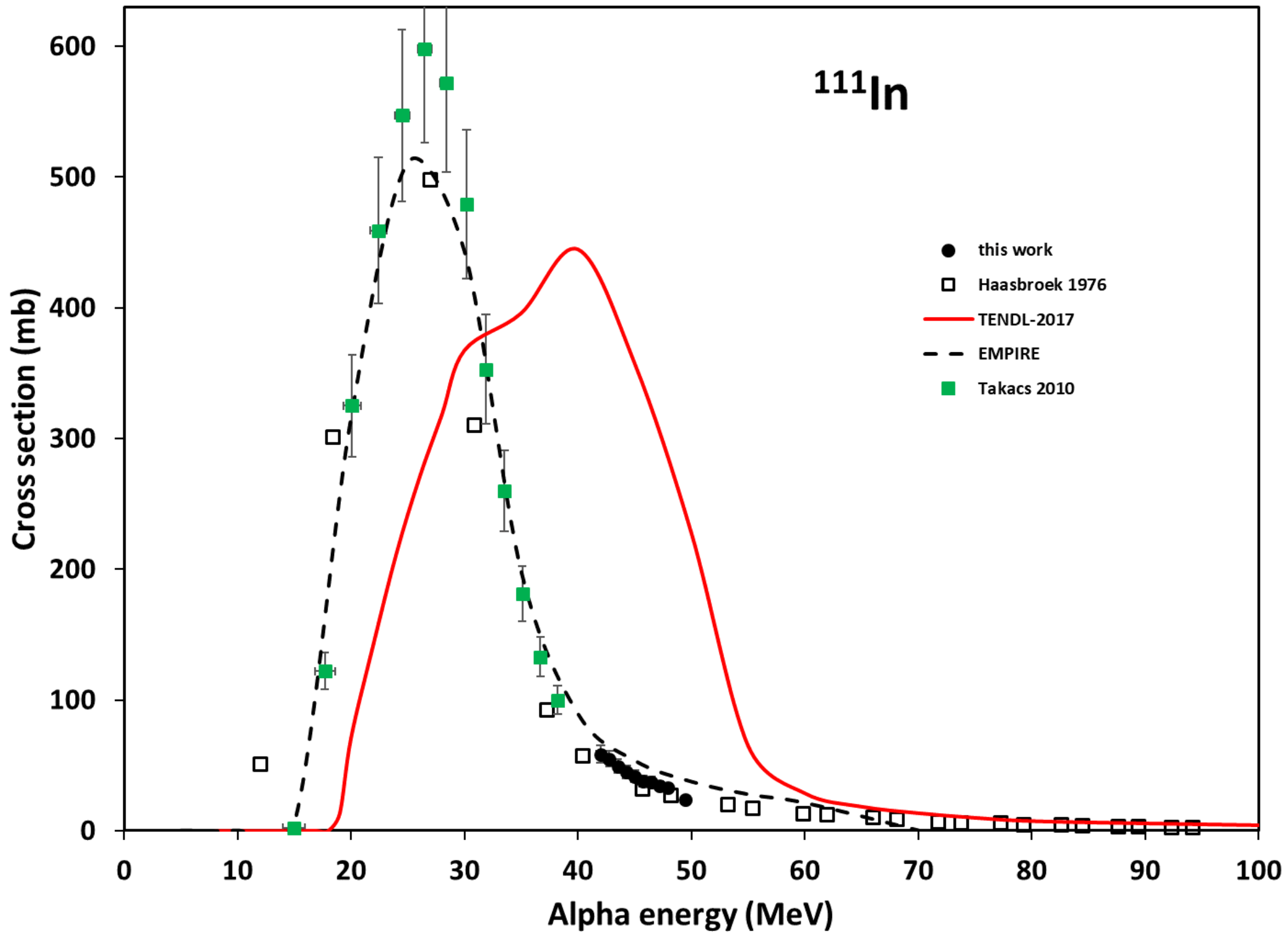

Fig. 2 Measured excitation function of the $^{nat}Ag(\alpha,x)^{111}In$ nuclear reaction compared with the previous results from the literature and with the results of theoretical model code calculations

From Fig 2. it is seen that our new data are in excellent agreement with the previous results of Haasbroek et al. [6] and Takács et al. [10] in the overlapping energy range and also with the prediction of the EMPIRE 3.2 calculations. EMPIRE also follow the results of the previous experiments in the lower and upper energy regions. TENDL-2017 does not predict the maximum around 28 MeV correctly even produces two local maxima. The values given by TENDL-2017 are acceptable only above 65 MeV.

### 4.2 $^{nat}Ag(\alpha,x)^{110g}In$ nuclear reaction

This radioisotope is produced with 1 and 3 neutron emissions from the $^{107}Ag$ and $^{109}Ag$ target isotopes, respectively. It has a relatively short half-life (4.9 h), so only the earliest measured spectra could be used for the final evaluation. The decay of $^{110g}In$ is followed by emission of several strong gamma-lines, but because of interferences the 937.48 keV peak was used for the final calculations. While both the ground-state and the meta-stable state decay independently to $^{110}Cd$ the cross sections of both states could be determined independently.

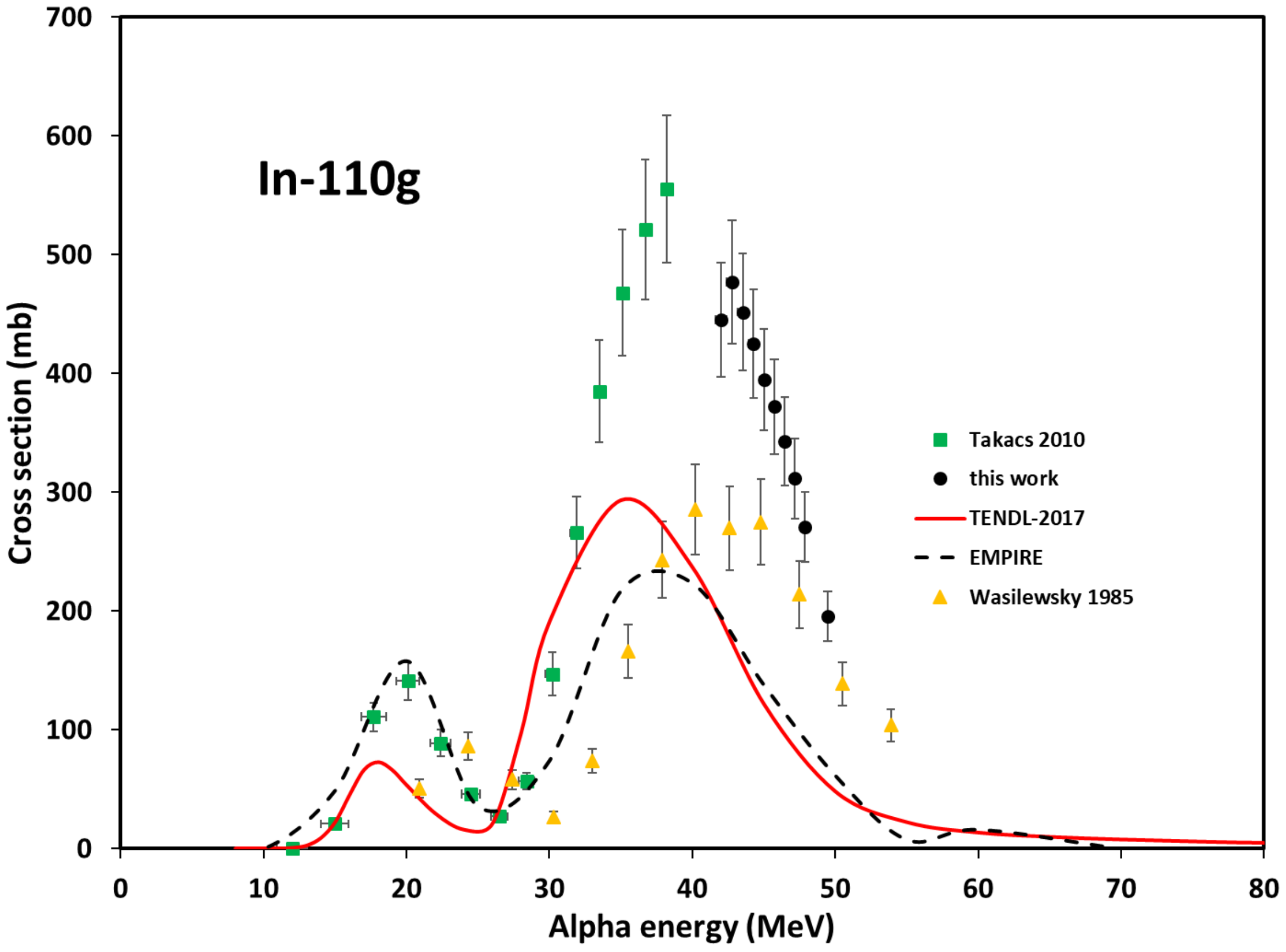

Fig. 3 Measured excitation function of the $^{nat}Ag(\alpha,x)^{110g}In$ nuclear reaction compared with the previous results from the literature and with the results of theoretical model code calculations

$^{110g}In$ has several strong gamma-lines, but utmost care should have been taken because most of them interfere with the excited state of $^{110}In$ and/or with other produced radioisotopes.

From Fig. 3 it is seen that our new data give a good continuation to the previous results of Takács et al. [10] at the high energy part of the local maximum on the excitation function. The data of Wasilewsky [12] was recalculated for absolute cross section. The trend is similar except the large energy shift and lower values below 50 MeV. Both TENDL-2017 and EMPIRE 3.2 give similar trend, EMPIRE describes the low energy maximum well both in position and in amplitude. The higher energy local maximum is a little bit shifted by TENDL-2017, while the position prediction of EMPIRE 3.2 is better. Both model codes underestimate the experimental values around the higher energy maximum.

### 4.3 $^{nat}Ag(\alpha,x)^{110m}In$ nuclear reaction

The $^{110}$In radionuclide has a meta-stable state with an only 69.1 min half-life. It means that only data from the earliest measured spectra could be used for evaluation. The earliest measurements began about 10 hours after the end of bombardment and only the last Ag foil (lowest energy) could be measured. In the time of the next Ag measurement there was no measurable $^{110m}$In activity in the samples. The only strong and measurable gamma-line is the 657.65 keV line, but unfortunately the ground-state and $^{110m}$Ag have this same line. The contribution of the $^{110m}$In meta-stable state could be easily separated by using the independent gamma-lines of its ground-state. In the case of the $^{110m}$Ag isotope the situation was a bit more complicated, because it had no peaks with reasonable statistics in the earliest measured series, due to its long half-life and the short measuring time. The result is presented in Fig. 4. Our new point is closer to the results of Patel et al. [8]. The theoretical model codes estimate the first two maxima almost correctly, but give different values above 25 MeV.

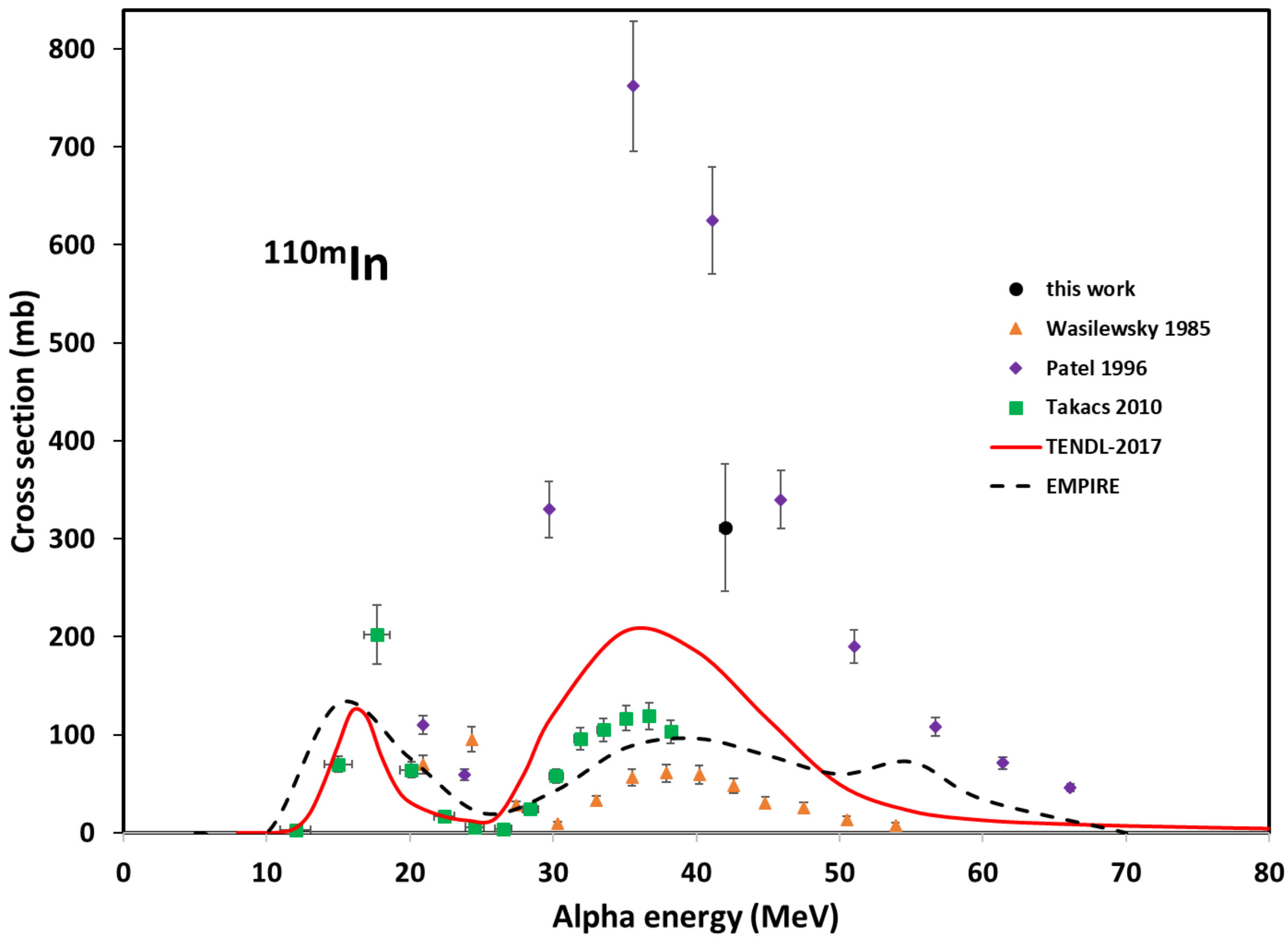

Fig. 4 Measured excitation function of the $^{nat}Ag(\alpha,x)^{110m}In$ nuclear reaction compared with the previous results from the literature and with the results of theoretical model code calculations

### 4.4 $^{nat}Ag(\alpha,x)^{109g}In$ nuclear reaction

The $^{109g}In$ radionuclide is produced by direct reactions from the $^{107}Ag$ and $^{109}Ag$ target isotopes with 2 and 4 neutrons emission, respectively. Because of its relatively short half-life (4.159 h) reasonable results could only be assessed from the first measurement series. Due to its short-lived (1.34 min) meta-stable state, which completely decays into the ground-state, the measured cross section is cumulative. The 203.3 MeV gamma-line, which was used for the evaluation, is independent. The results are presented in Fig. 5.

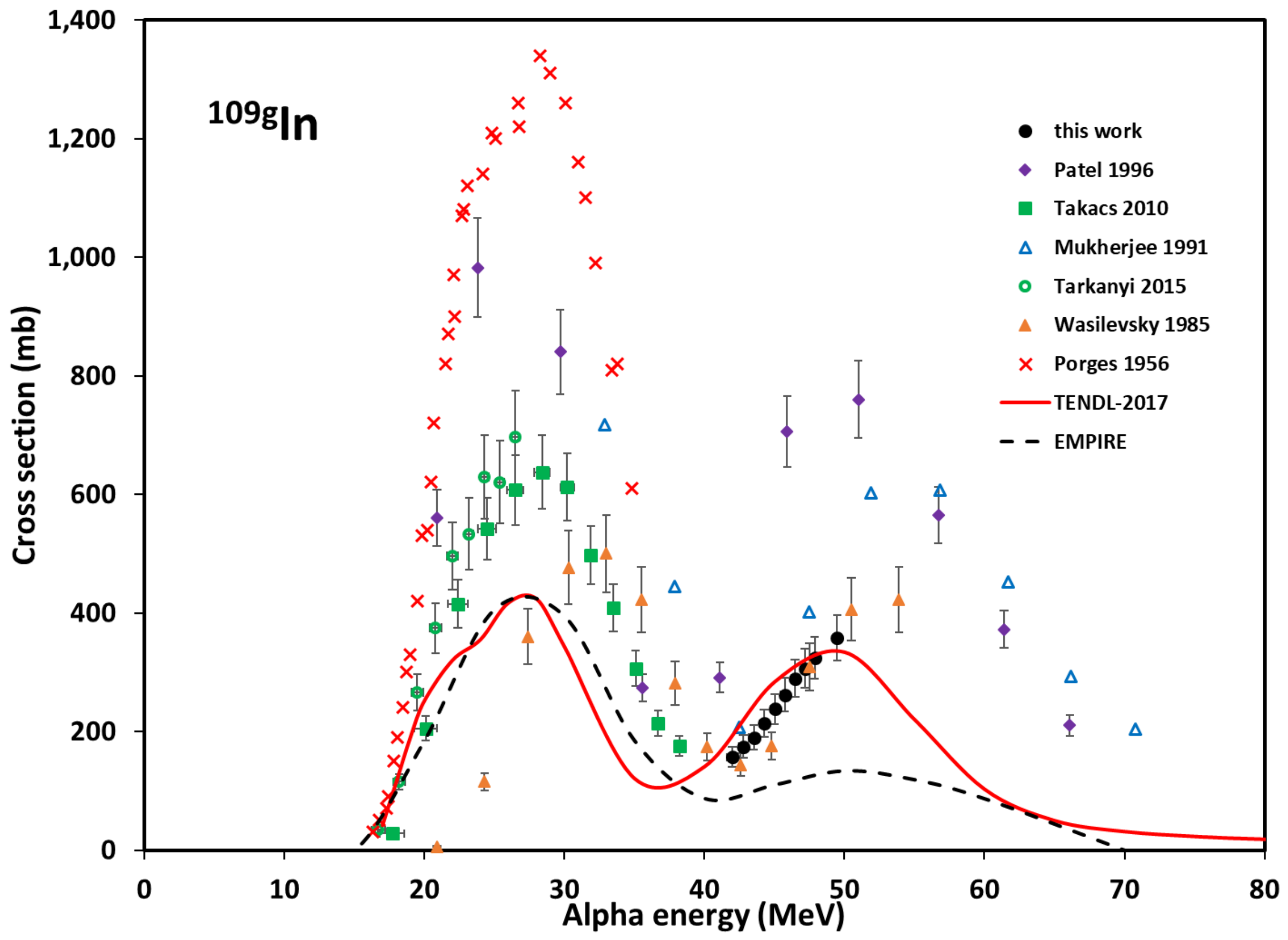

Fig. 5 Measured excitation function of the $^{nat}Ag(\alpha,x)^{109g}In$ nuclear reaction compared with the previous results from the literature and with the results of theoretical model code calculations

From Fig. 5 it is seen that our new results rather support the previous data of Takács et al. [10] giving a good continuation to their previous data. All the other data show large discrepancies, only the data of Wasilewsky agree well with our new results in spite of their large energy shift in the lower energy region. The agreement with the TENDL-2017 prediction is also very good. The both theoretical model codes give similar trend and describe both local maxima correctly, but EMPIRE 3.2 underestimates the cross sections in the energy region above 40 MeV.

### 4.5 $^{nat}$Ag($\alpha$,x)$^{108g}$In nuclear reaction

The $^{108g}$In radionuclide is produced by direct reactions from the $^{107}$Ag and $^{109}$Ag target isotopes with 3 and 5 neutrons emission, respectively. It has a relatively short half-life (58 min) so the results for this reaction were deduced only from the first spectrum of the first series, which means that we could only deduce a single experimental point (see Fig. 6). The even shorter-lived isomeric state $^{108m}$In (38,6 min, which could not be assessed from this experiment) decays completely into the stable $^{108}$Cd, so cumulative effect and/or interfering radiation do not take place. The results are presented in Fig. 6.

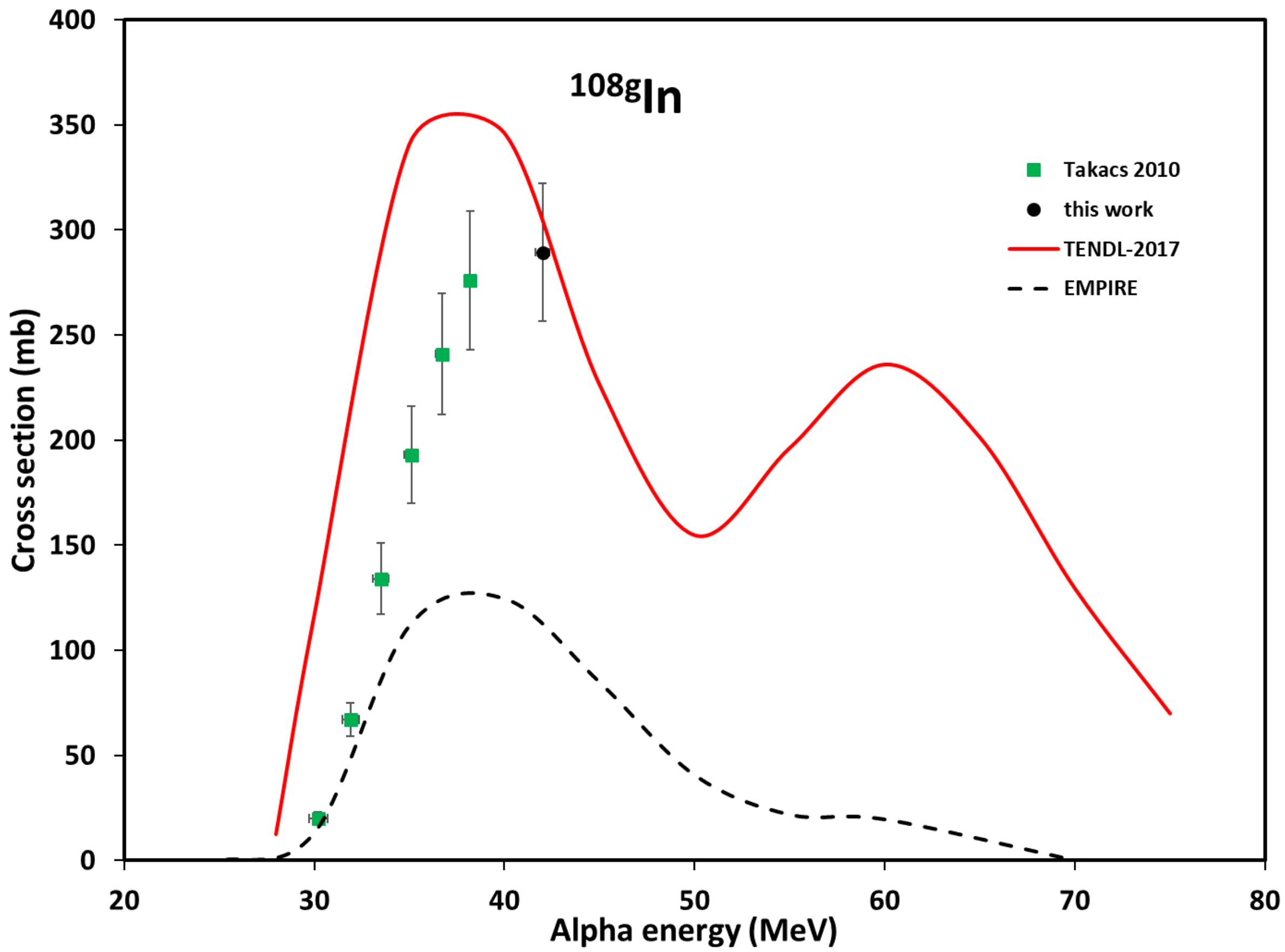

Fig. 6 Measured excitation function of the $^{nat}Ag(\alpha,x)^{108g}In$ nuclear reaction compared with the previous results from the literature and with the results of theoretical model code calculations

From Fig. 6 it is seen that the newly measured single data point gives a good continuation to the previous data of Takács et al. [10]. The two theoretical model codes give similar trends with different values, and the results from TENDL-2017 seem to be a better approximation. It is also seen from the results of the model calculations that mainly the nuclear reactions on $^{107}Ag$ dominate for the production of the $^{108}In$ radioisotope.

### 4.6 $^{nat}Ag(\alpha,x)^{111g}Ag$ nuclear reaction

The radioisotope $^{111}Ag$ has a very short-lived isomeric state (65 s) and a relatively longer-lived (7.45 d) ground-state. The isomeric state decays completely into the ground-state and because of its very short half-life all measurements series gave cumulative cross sections for the $^{111}Ag$ radioisotope. $^{111}Ag$ is produced exclusively from the $^{109}Ag$ target isotope with two protons emission. Taking into account the longer half-life the best (but not very good) statistics was resulted in from the measurement series with longer cooling and measuring times. The 342.13 keV independent gamma-line was used for the evaluation. The results are presented in Fig. 7.

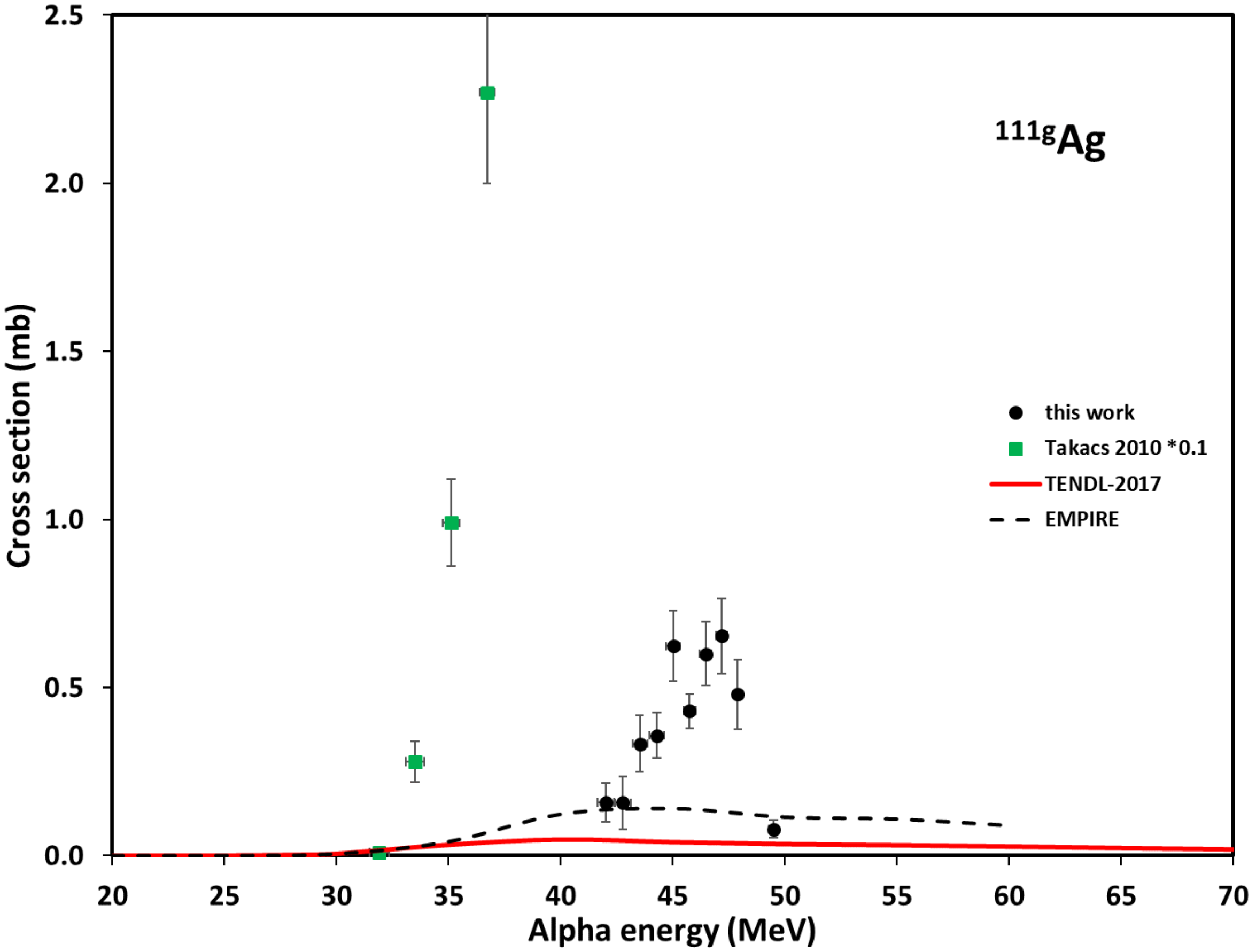

Fig. 7 Measured excitation function of the $^{nat}Ag(\alpha,x)^{111g}Ag$ nuclear reaction compared with the previous results from the literature and with the results of theoretical model code calculations

From Fig. 7 it is seen that our new results are completely different from the previously measured data. Both the theoretical model calculations cannot reproduce our new results, though EMPIRE gives closer values to our new experimental results.

#### 4.7 $^{nat}Ag(\alpha,x)^{110m}Ag$ nuclear reaction

The $^{110}$Ag has a long-lived (249.76 d) isomeric state and a short-lived ground state (24.6 s). The ground state in this experiment could not be measured. The results were deduced from the measurement series with the longest cooling times and longest measuring times. The results are presented in Fig. 8.

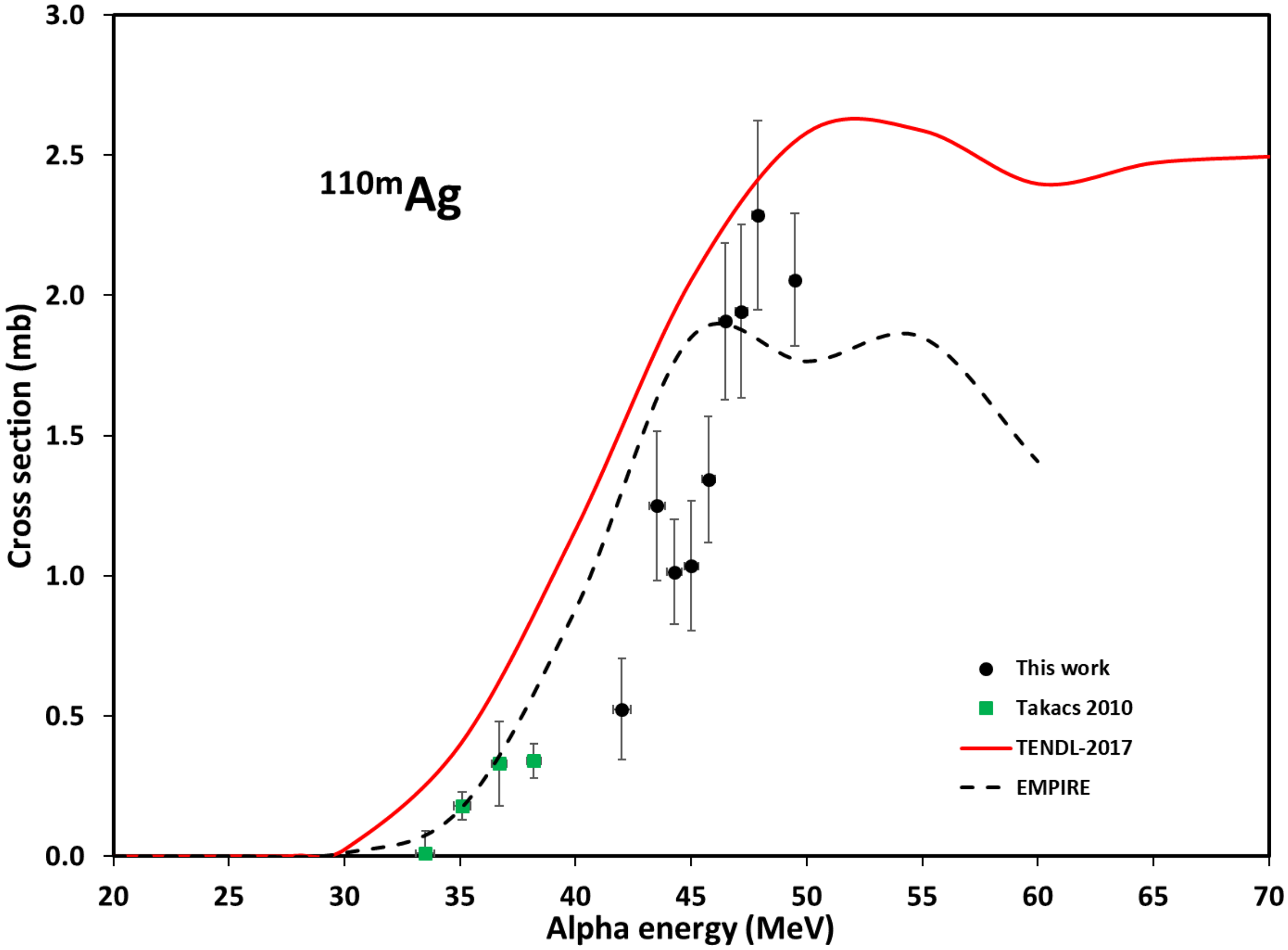

Fig. 8 Measured excitation function of the $^{nat}$Ag($\alpha$,x)$^{110m}$Ag nuclear reaction compared with the previous results from the literature and with the results of theoretical model code calculations

From Fig. 8 it is seen that our new data give a good continuation to the previous results of Takács et al. [10]. The EMPIRE 3.2 prediction is in reasonably good agreement with our new data and the previous results. The results from the TENDL-2017 on-line library give a bit higher values up to 45 MeV and a different trend above this energy.

### 4.8 $^{nat}Ag(\alpha,x)^{106m}Ag$ nuclear reaction

The radioisotope $^{106}Ag$ can be produced from both stable target isotopes $^{107}Ag$ and $^{109}Ag$ by emissions of 2 protons, and 3 and 5 neutrons, respectively. The ground state has a too short half-life (23.96 min) to be included in our evaluation and anyway has no proper gamma-line. The excited state (8.28 d) has several relatively strong gamma-lines, so those having good statistics were involved into the evaluation. The results are shown in Fig. 9.

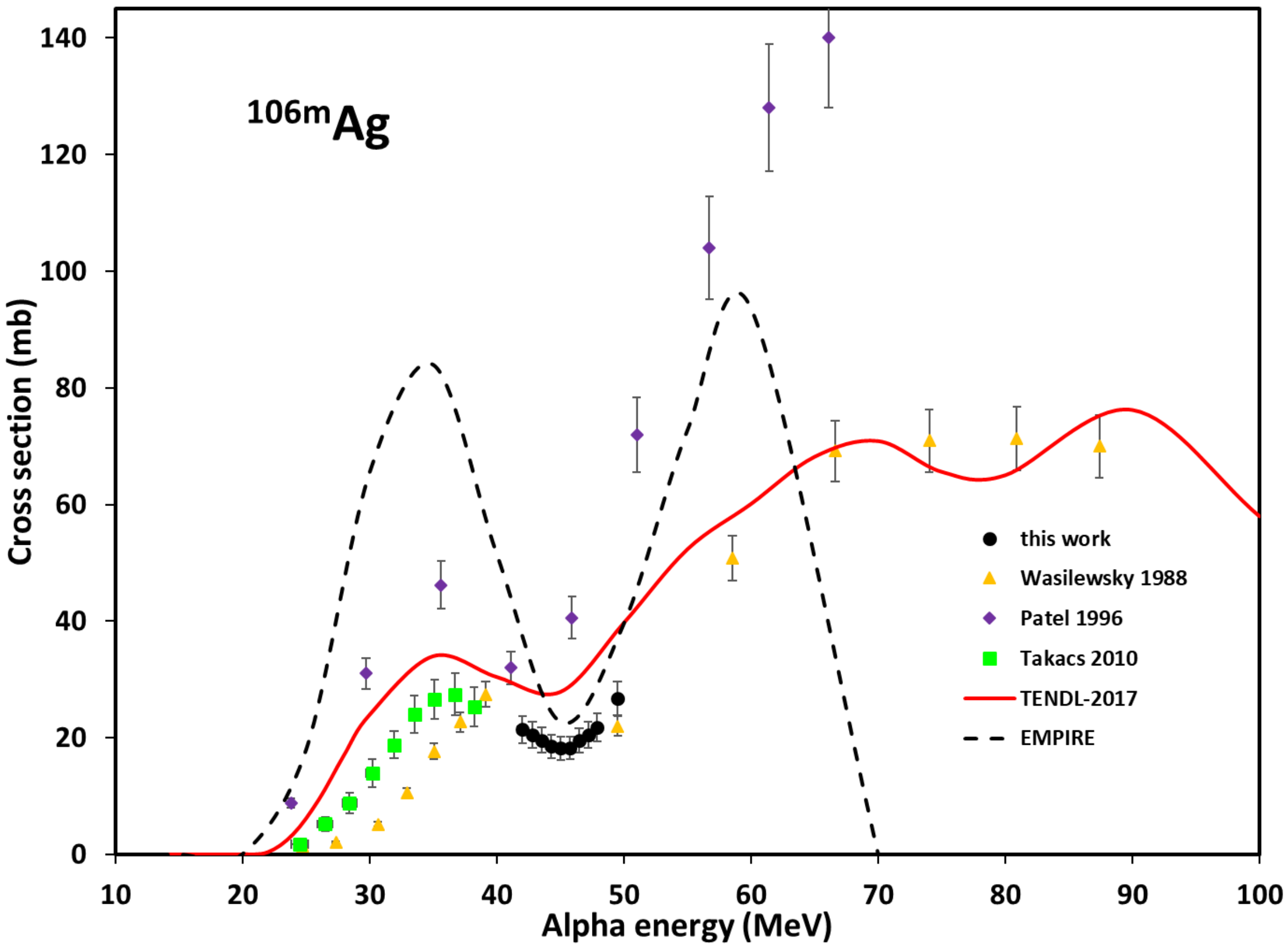

Fig. 9 Measured excitation function of the $^{nat}Ag(\alpha,x)^{106m}Ag$ nuclear reaction compared with the previous results from the literature and with the results of theoretical model code calculations

The new data in Fig. 9 show a good continuation to the previous results of Takács et al. [10], lower than the data of Patel et al. [8] and show a good agreement with the data of Wasilewsky et al [12], in spite of their energy shift. Both TENDL-2017 and EMPIRE 3.2 give completely different trends and values. Up to 50 MeV the TENDL-2017 estimation seems to be better with a small overestimation.

### 4.9 $^{nat}Ag(\alpha,x)^{105g}Ag$ nuclear reaction

The $^{105}Ag$ has two isomeric states, a short-lived (7.23 min) metastable state, which could not be assessed in this work and a longer-lived (41.29 d) ground state. The higher energy isomer state decays 100% into the ground-state.

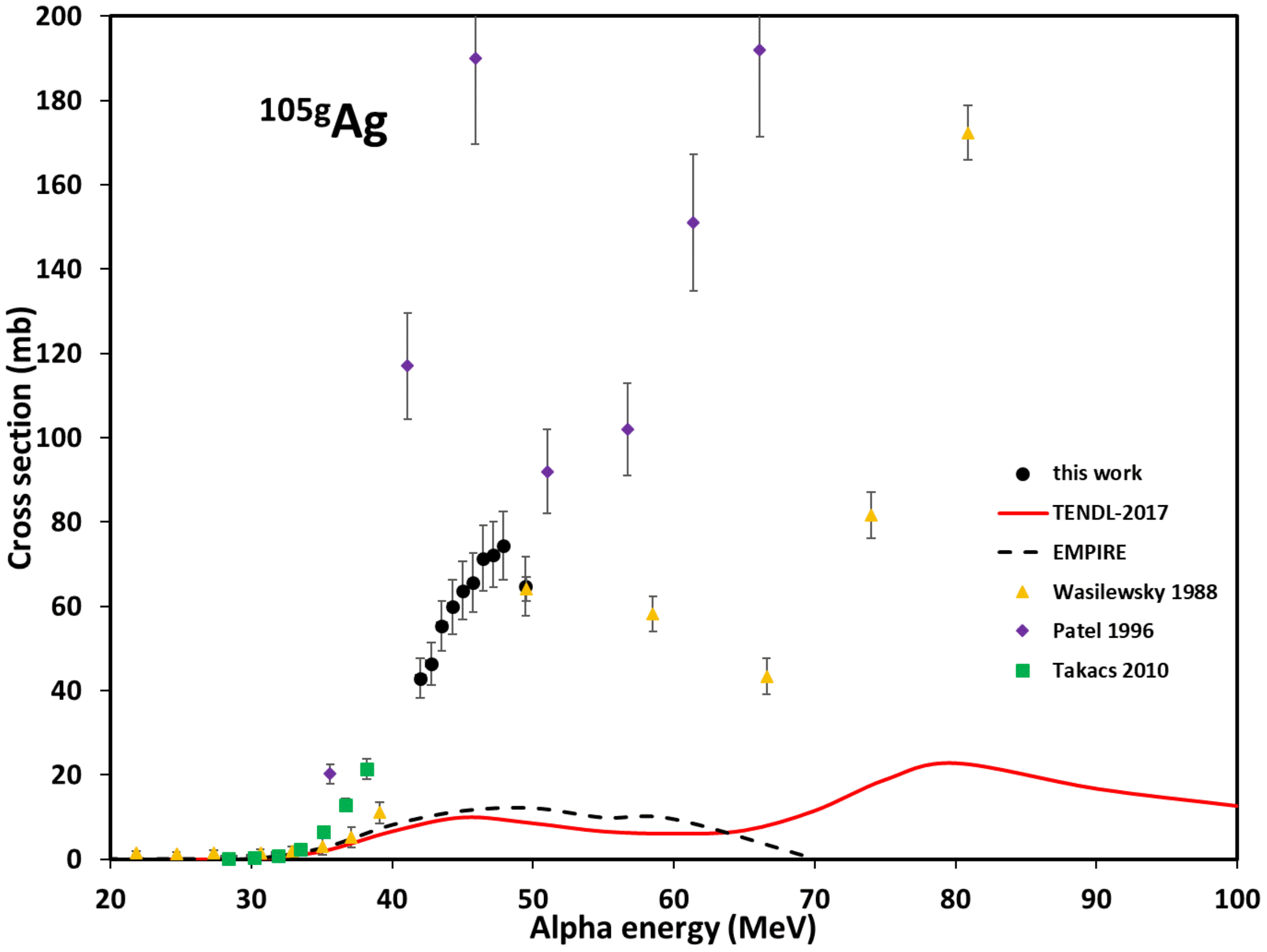

Fig. 10 Measured excitation function of the $^{nat}Ag(\alpha,x)^{105g}Ag$ nuclear reaction compared with the previous results from the literature and with the results of theoretical model code calculations

From Fig. 10 it is seen that the newly deduced data give a good continuation to the previous results of Takács et al. [10]. Data points of Patel et al. [8] are scattered. The previous results of Wasilewsky [12] show partial agreement, but the energy shift is still visible. Both the theoretical model code predictions strongly underestimate the experiments above 35 MeV, though they give the same trend and values up to 65 MeV.

### 4.10 $^{nat}Ag(\alpha,x)^{109}Cd$ nuclear reaction

$^{109}Cd$ can be produced from both the stable isotopes of Ag target elements with the emission of one proton and several neutrons. The $^{109}Cd$ radionuclide has a long half-life (461.9 d), and the co-produced $^{109}In$ decays into it, so reliable results could be deduced from the long measurements with long cooling time. Only its low-intensity gamma peak at 88.03 keV could be measured in this work. The cumulative results are presented in Fig. 11.

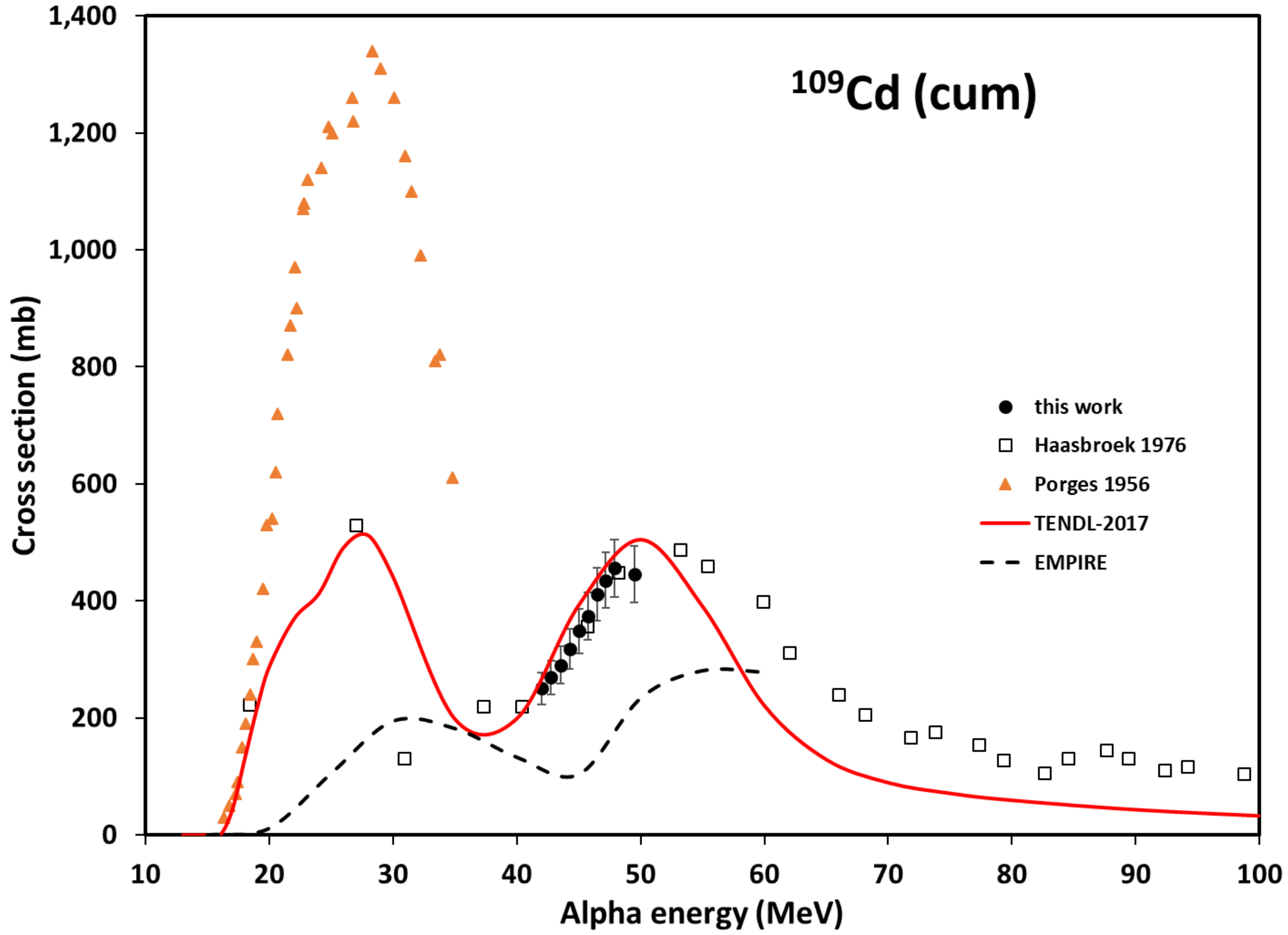

Fig. 11 Measured excitation function of the $^{nat}Ag(\alpha,x)^{109}Cd$ nuclear reaction compared with the previous results from the literature and with the results of theoretical model code calculations

From Fig. 11 it is seen that our new data are in excellent agreement with the previous results of Haasbroek [6] and also with the prediction of the TENDL-2017 on-line data library. The results of Porges [9] are not comparable, because they are in the energy range below 35 MeV. Both the theoretical model codes give similar trends describing the contributions on the two target isotopes. EMPIRE seems to be energy-shifted and the predicted amplitudes of TENDL-2017 are better at least in the 38-55 MeV energy range.

**Table 2** Experimental cross sections of the $^{nat}Ag(\alpha,x)^{111,110g,110m,109g,108g}In$ nuclear reactions

| Alpha energy | | Cross section (mb) | | | | | | | | | |
|---|---|---|---|---|---|---|---|---|---|---|---|
| MeV | | $^{111}In$ | | $^{110g}In$ | | $^{110m}In$ | | $^{109g}In$ | | $^{108g}In$ | |
| E | ΔE | σ | Δσ | σ | Δσ | σ | Δσ | σ | Δσ | σ | Δσ |
| 49.49 | 0.20 | 24.06 | 2.60 | 195.52 | 21.16 | | | 357.83 | 38.71 | | |
| 47.89 | 0.24 | 32.91 | 3.56 | 270.67 | 29.37 | | | 323.90 | 35.07 | | |
| 47.18 | 0.25 | 34.28 | 3.71 | 311.59 | 33.77 | | | 306.29 | 33.18 | | |
| 46.47 | 0.27 | 37.28 | 4.03 | 342.95 | 37.16 | | | 289.45 | 31.36 | | |
| 45.75 | 0.29 | 38.10 | 4.12 | 372.00 | 40.28 | | | 262.03 | 28.38 | | |
| 45.02 | 0.30 | 41.48 | 4.49 | 394.68 | 42.74 | | | 237.73 | 25.75 | | |
| 44.28 | 0.32 | 44.74 | 4.84 | 425.05 | 46.03 | | | 213.25 | 23.11 | | |
| 43.53 | 0.34 | 49.41 | 5.34 | 451.76 | 48.90 | | | 189.55 | 20.53 | | |
| 42.77 | 0.35 | 55.07 | 5.96 | 477.04 | 51.64 | | | 174.56 | 18.92 | | |
| 42.01 | 0.37 | 58.52 | 6.33 | 445.35 | 48.21 | 311.15 | 65.00 | 156.57 | 16.94 | 289.35 | 32.68 |

**Table 3** Experimental cross sections of the $^{nat}Ag(\alpha,x)^{111g,110m,106m,105g}Ag,^{109}Cd$ nuclear reactions

| Alpha energy | | Cross section (mb) | | | | | | | | | |
|---|---|---|---|---|---|---|---|---|---|---|---|
| MeV | | $^{111}Ag$ | | $^{11om}Ag$ | | $^{106m}Ag$ | | $^{105g}Ag$ | | $^{109}Cd$ | |
| E | ΔE | σ | Δσ | σ | Δσ | σ | Δσ | σ | Δσ | σ | Δσ |
| 49.49 | 0.20 | 0.08 | 0.03 | 2.06 | 0.24 | 26.76 | 2.90 | 64.83 | 7.01 | 445.62 | 48.21 |
| 47.89 | 0.24 | 0.48 | 0.10 | 2.29 | 0.34 | 21.80 | 2.37 | 74.36 | 8.05 | 455.86 | 49.41 |
| 47.18 | 0.25 | 0.65 | 0.11 | 1.94 | 0.31 | 20.44 | 2.22 | 72.30 | 7.83 | 435.09 | 47.16 |
| 46.47 | 0.27 | 0.60 | 0.10 | 1.91 | 0.28 | 19.51 | 2.12 | 71.39 | 7.73 | 411.08 | 44.54 |
| 45.75 | 0.29 | 0.43 | 0.05 | 1.34 | 0.22 | 18.23 | 1.99 | 65.63 | 7.11 | 373.44 | 40.47 |
| 45.02 | 0.30 | 0.62 | 0.10 | 1.04 | 0.23 | 18.20 | 1.98 | 63.77 | 6.91 | 348.02 | 37.69 |
| 44.28 | 0.32 | 0.36 | 0.07 | 1.01 | 0.19 | 18.52 | 2.01 | 59.91 | 6.48 | 317.39 | 34.35 |
| 43.53 | 0.34 | 0.33 | 0.08 | 1.25 | 0.27 | 19.58 | 2.13 | 55.32 | 5.99 | 290.29 | 31.54 |
| 42.77 | 0.35 | 0.16 | 0.08 | | | 20.53 | 2.23 | 46.44 | 5.03 | 268.70 | 29.25 |
| 42.01 | 0.37 | 0.16 | 0.06 | 0.52 | 0.18 | 21.40 | 2.33 | 42.96 | 4.66 | 250.41 | 27.18 |

## 5 Physical yield

Integral physical yields have been calculated from the measured excitation functions for the investigated nuclear reactions. The experimental literature data were used to complete the low energy part of the excitation function, if they were available, in order to cover the whole energy region in the yield calculation. Searching the literature only one previous work was found [26] with experimental thick target yield measurements on natural silver.

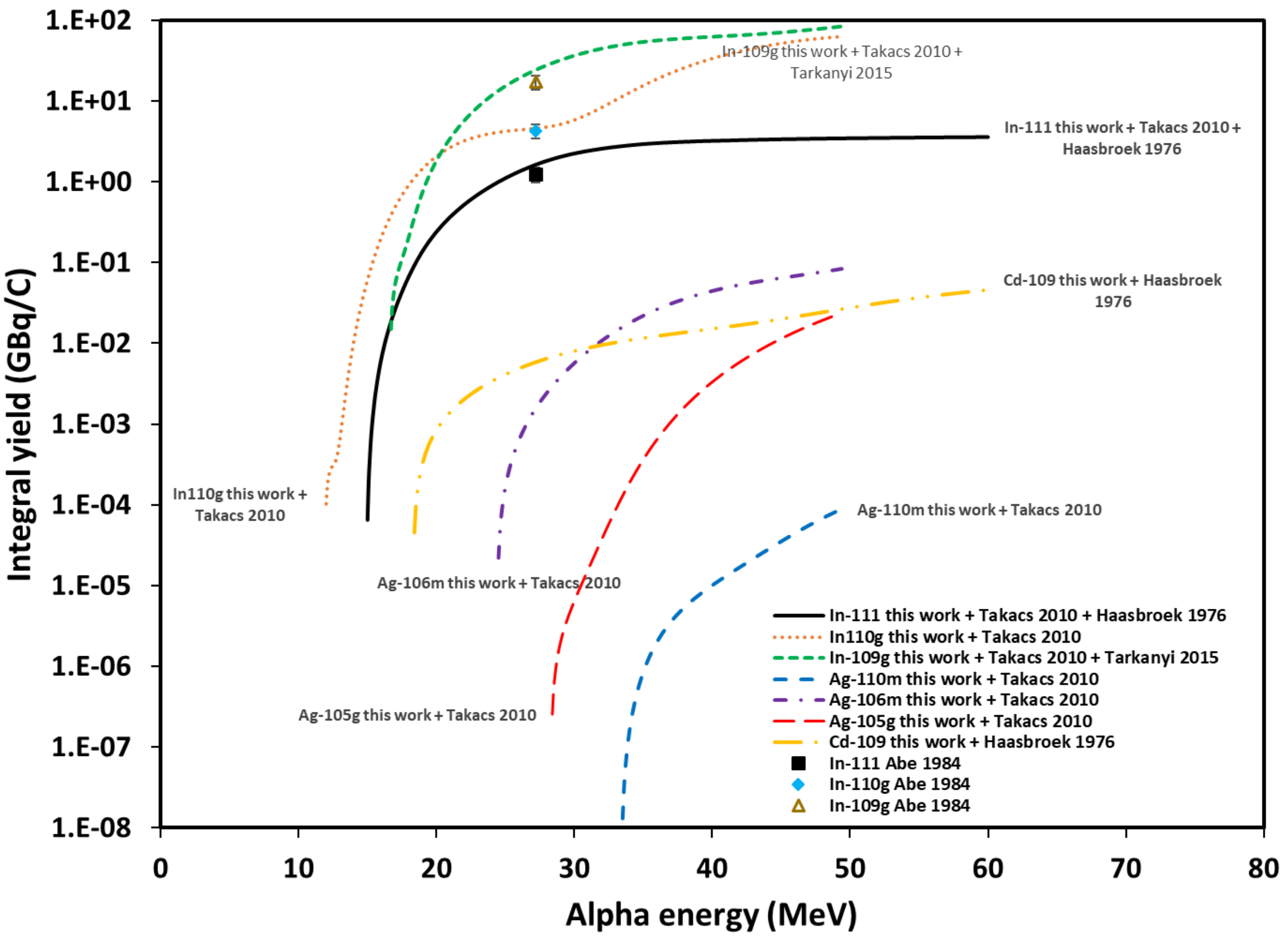

Fig 12 Calculated physical yields from selected $\alpha$-particle induced nuclear reactions on natural silver compared with the literature data

From Fig. 12 it is seen that the yields cover 10 orders of magnitudes in different energy ranges. The highest production yields are found for the two reactions to produce the medically interesting $^{109g}$In and $^{110g}$In isotopes. Both isotopes can be produced in considerable amount (above 1GBq/C) using proper energy range (above 20 MeV) cyclotrons. Considering the deduced thick target yield for three radioisotopes ($^{110g}$In, $^{109g}$In and $^{111}$In), good agreements were found with the previously published results by Abe et al. [26].

## 6 Industrial applications

Beside the medical applications some of the produced radioisotopes can also be used in the industry as radioactive tracers. The selection criteria are the proper (enough long) half-life, proper (medium energy and high intensity) gamma-lines, and good production parameters. The first two criteria limit the selection for the Ag radioisotopes, except $^{111}$Ag which has only weak gamma lines. The $^{106m}$Ag has the highest production yield and the production energy window still can be reached by medium energy accelerators. Its half-life allows tracing of relatively quick processes within a month application time. On the other hand, the shorter half-life ensures low environmental load.

The most interesting application of radioisotopes in tracer technology is TLA (Thin Layer Activation) [27, 28]. With $^{106m}$Ag it is possible to perform TLA investigations to measure wear, corrosion or erosion rate of parts containing silver, within a period of about one month. As an example four activity distributions are presented in Fig. 13, with 4 different parameter sets. All the four activity-depth distribution provide almost a constant activity for a certain depth measured from the surface down (see Table 4).

Table 4 Parameters of activity distributions in Fig. 13. (EOB=End of Bombardment)

| | | | | |
|---|---|---|---|---|
| Bombarding energy (MeV) | 37.6 | 37.6 | 37.6 | 37.6 |
| Irradiation angle | 90° | 90° | 15° | 15° |
| Cooling time (day) | 0(EOB) | 10 | 0(EOB) | 10 |
| Irradiation time (h) | 1 | 1 | 1 | 1 |
| Beam current (μA) | 2 | 2 | 2 | 2 |
| Total activity (kBq) | 238.24 | 103.15 | 238.24 | 103.15 |
| Activity  of the constant layer (kBq) | 58.35 | 25.26 | 58.35 | 25.26 |
| Specific activity at the surface (kBq/μm) | 3.33 | 1.44 | 12.88 | 5.58 |
| Thickness of the quasi-constant layer (μm) | 17.50 | 17.50 | 4.53 | 4.53 |

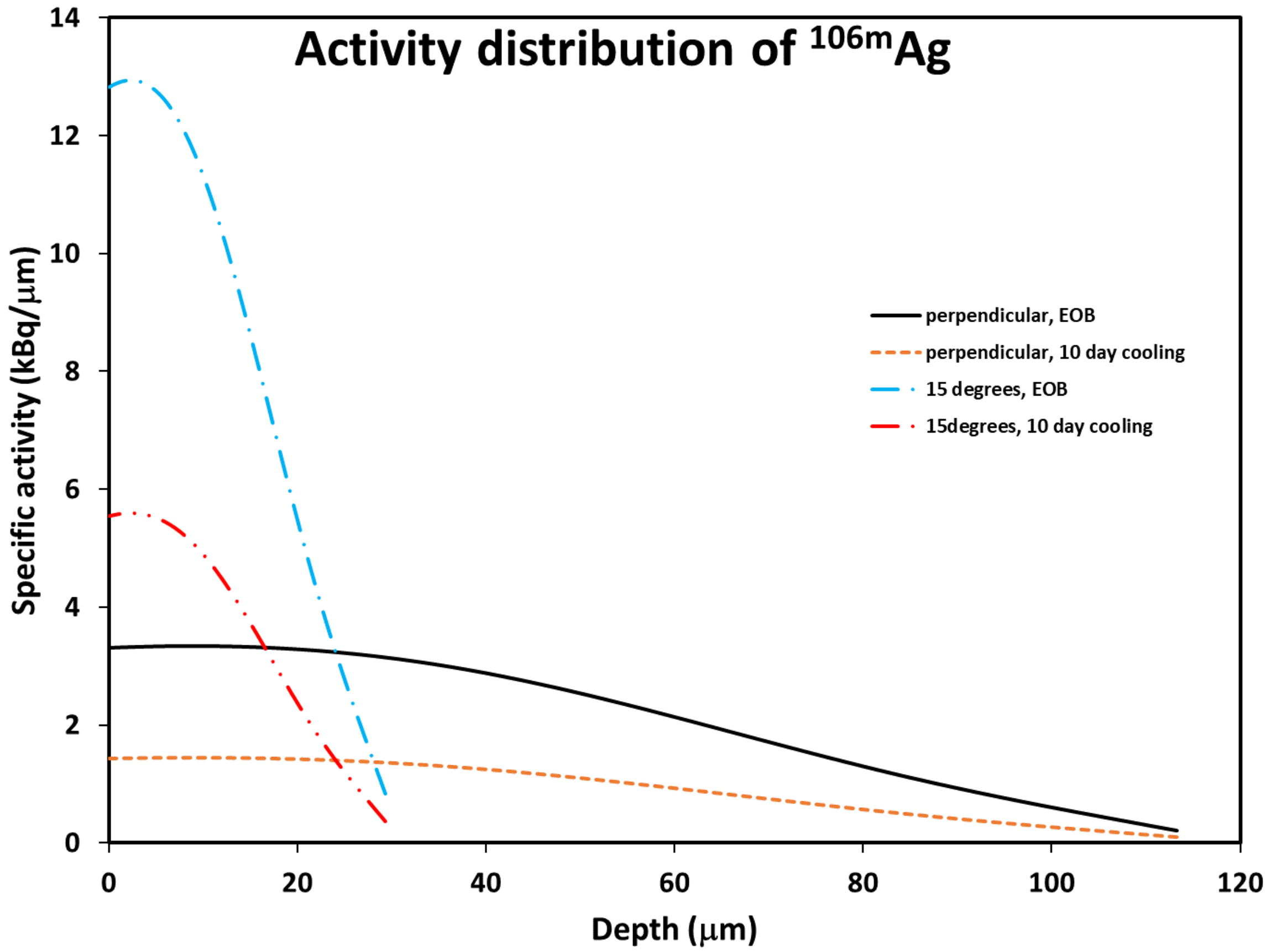

Fig. 13 Activity distribution of the $^{106m}Ag$ radioisotope irradiated and measured according to the parameters in Table 4.

From Fig. 13 and Table 4 it is seen that already by 1 hour irradiation reasonable specific activity can be produced in the surface layer for wear measurements, not only at the end of the irradiation (EOB) but even after 10 days cooling time. The bombarding energy (37.6 MeV) was chosen in such a way that in the first several μm-s from the irradiated surface the activity is almost homogeneous (and the specific activity is almost constant) for a certain depth. This type of activity depth distribution can be reached for those nuclear reactions, which has an excitation function with a local maximum in the available energy range. By changing the irradiation angle the depth profile of the produced activity (and the thickness of the constant specific activity layer) can be controlled together with the specific activity. The total and the specific activity can be multiplied by increasing the beam current. A longer irradiation also provides higher activity but the irradiation time vs. activity function is not a linear one. The produced total activity should be below the FHL (Free Handling Limit =1000 kBq for $^{106m}Ag$)[28] level, if the industrial laboratory or site has no license to handle and store radioactive materials, which is the case according to Table 4. For a licensed laboratory higher activity can be produced (with additional transport problems).

## 7 Conclusions

Excitation functions for the $^{nat}Ag(\alpha,x)^{111,110m,110g,109g,108g}In$,$^{111,110m,106m,105g}Ag$,$^{109}Cd$ were measured in the energy range of 40-50 MeV. The newly determined cross section data help to clarify the problems between the previous literature results. Cross section deduced for production of $^{110g}In$, $^{109g}In$, $^{108g}In$, $^{111}Ag$ and $^{110g}Ag$ in most cases show a good continuation of the eventually existing literature data in a lower energy region. In the case of $^{111}In$ and $^{109}Cd$ the agreement with the previous literature data is excellent.

The results of the theoretical nuclear reaction model codes are not systematic and give only partly good estimations for several reactions. There are reactions, for which both (EMPIRE and TENDL) completely fail.

From the measured cross sections thick target physical yield curves were calculated. The excitation functions for these calculations were constructed by using our new results combined with data from the literature. The literature values agree well with our results.

Among the possible industrial applications, the TLA method was demonstrated by using the $^{106m}Ag$ the best radioisotope for this purpose. It has been proved that by using $^{106m}Ag$ as tracer, the wear measurement with actual parameters can be performed.

***Acknowledgements***

This work was performed at the RI Beam Factory operated by the RIKEN Nishina Center and CNS, University of Tokyo. The authors express their gratitude to the crew of the AVF cyclotron for their invaluable assistance in the course of the experiment. The work of Masayuki Aikawa was supported by JSPS KAKENHI Grant Number 17K07004.

**Figure captions**

1. Re-measured cross section of the $^{nat}Ti(\alpha,x)^{51}Cr$ monitor reaction compared with the recommended data
2. Measured excitation function of the $^{nat}Ag(\alpha,x)^{111}In$ nuclear reaction compared with the previous results from the literature and with the results of theoretical model code calculations
3. Measured excitation function of the $^{nat}Ag(\alpha,x)^{110g}In$ nuclear reaction compared with the previous results from the literature and with the results of theoretical model code calculations
4. Measured excitation function of the $^{nat}Ag(\alpha,x)^{110m}In$ nuclear reaction compared with the previous results from the literature and with the results of theoretical model code calculations
5. Measured excitation function of the $^{nat}Ag(\alpha,x)^{109g}In$ nuclear reaction compared with the previous results from the literature and with the results of theoretical model code calculations
6. Measured excitation function of the $^{nat}Ag(\alpha,x)^{108g}In$ nuclear reaction compared with the previous results from the literature and with the results of theoretical model code calculations
7. Measured excitation function of the $^{nat}Ag(\alpha,x)^{111g}Ag$ nuclear reaction compared with the previous results from the literature and with the results of theoretical model code calculations
8. Measured excitation function of the $^{nat}Ag(\alpha,x)^{110m}Ag$ nuclear reaction compared with the previous results from the literature and with the results of theoretical model code calculations
9. Measured excitation function of the $^{nat}Ag(\alpha,x)^{106m}Ag$ nuclear reaction compared with the previous results from the literature and with the results of theoretical model code calculations

10. Measured excitation function of the $^{nat}Ag(\alpha,x)^{105g}Ag$ nuclear reaction compared with the previous results from the literature and with the results of theoretical model code calculations
11. Measured excitation function of the $^{nat}Ag(\alpha,x)^{109}Cd$ nuclear reaction compared with the previous results from the literature and with the results of theoretical model code calculations
12. Calculated physical yields from selected $\alpha$-particle induced nuclear reactions on natural silver compared with the literature data
13. Activity distribution of the $^{106m}Ag$ radioisotope irradiated and measured according to the parameters in Table 4.

## References

[1] M. Roca, E.F.J. de Vries, F. Jamar, O. Israel, A. Signore, Guidelines for the labelling of leucocytes with (111)In-oxine, Eur J Nucl Med Mol I, 37 (2010) 835-841.

[2] P.V. Kulkarni, Single-photon emitting radiotracers produced by cyclotrons for myocardial imaging, Nucl. Instrum. Methods Phys. Res., Sect. B, 40-41 (1989) 1114-1117.

[3] M. Sadeghi, M. Mirzaee, Z. Gholamzadeh, A. Karimian, F.B. Novin, Targetry and radiochemistry for no-carrier-added production of Cd-109, Radiochim. Acta, 97 (2009) 113-116.

[4] M. Mirzaii, M. Sadeghi, Z. Gholamzadeh, Targetry for cyclotron production of no-carrier-added cadmium-109 from Ag-nat(p,n)Cd-109 reaction, Iran J Radiat Res, 6 (2009) 201-206.

[5] M. Fassbender, F.M. Nortier, D.R. Phillips, V.T. Hamilton, R.C. Heaton, D.J. Jamriska, J.J. Kitten, L.R. Pitt, L.L. Salazar, F.O. Valdez, E.J. Peterson, Some nuclear chemical aspects of medical generator nuclide production at the Los Alamos hot cell facility, Radiochim. Acta, 92 (2004) 237-243.

[6] F.J. Haasbroek, G.F. Burdzik, M. Cogneau, P. Wanet, Excitation Functions and Thick-Target Yields For Ga-67, Ge-68/Ga-68, Cd-109 and in-111 Induced in Natural Zinc and Silver by 100 MeV Alpha Particles in: Council f.Scient.and Indust.Res.,Pretoria,Repts., National Physical Research Lab., Pretoria, South Africa, Rep., Pretoria, 1976.

[7] S. Mukherjee, A.V.M. Rao, J.R. Rao, Preequilibrium Analysis of the Excitation-Functions of (Alpha,Xn) Reactions on Silver and Holmium, Nuovo Cimento A, 104 (1991) 863-874.

[8] H.B. Patel, M.S. Gadkari, B. Dave, N.L. Singh, S. Mukherjee, Analysis of the excitation function of alpha-particle-induced reactions on natural silver, Canadian Journal of Physics, 74 (1996) 618-625.

[9] K.G. Porges, Alpha excitation functions of silver and copper, Phys. Rev., 101 (1956) 225-230.

[10] S. Takács, A. Hermanne, F. Tárkányi, A. Ignatyuk, Cross-sections for alpha particle produced radionuclides on natural silver, Nucl. Instrum. Methods Phys. Res., Sect. B, 268 (2010) 2-12.

[11] F. Tárkányi, S. Takács, F. Ditrói, A. Hermanne, M. Baba, B.M.A. Mohsena, A.V. Ignatyuk, New cross section data and review of production routes of medically used $^{110m}$In, Nucl. Instrum. Methods Phys. Res., Sect. B, 351 (2015) 6-15.

[12] C. Wasilevsky, M.D. Vedoya, S.J. Nassiff, Isomer Yield Ratios and Cross-Sections for 110(4.9 H)in/110(69 Min)in and 108(58 Min)in/108(39.6 Min)in Produced by Alpha-Reactions on Silver, J. Radioanal. Nucl. Chem., 95 (1985) 29-44.

[13] T. Watanabe, M. Fujimaki, N. Fukunishi, H. Imao, O. Kamigaito, M. Kase, M. Komiyama, N. Sakamoto, K. Suda, M. Wakasugi, K. Yamada, Beam Energy and Longitudinal Beam Profile Measurement System at RIBF, in: 5th International Particle Accelerator Conference (IPAC2014), Jacow, Dresden, Germany, 2014, pp. 3566.

[14] Canberra, http://www.canberra.com/products/radiochemistry_lab/genie-2000-software.asp., in, 2000.

[15] G. Székely, Fgm - a flexible gamma-spectrum analysis program for a small computer, Comput. Phys. Commun., 34 (1985) 313-324.

[16] F. Tárkányi, S. Takács, K. Gul, A. Hermanne, M.G. Mustafa, M. Nortier, P. Oblozinsky, S.M. Qaim, B. Scholten, Y.N. Shubin, Z. Youxiang, Beam monitor reactions (Chapter 4). Charged particle cross-section database for medical radioisotope production: diagnostic radioisotopes and monitor reactions. , in: TECDOC 1211, IAEA, 2001, pp. 49.

[17] F. Tárkányi, F. Szelecsényi, S. Takács, Determination of effective bombarding energies and fluxes using improved stacked-foil technique, Acta Radiol., Suppl., 376 (1991) 72.

[18] International-Bureau-of-Weights-and-Measures, Guide to the expression of uncertainty in measurement, 1st ed., International Organization for Standardization, Genève, Switzerland, 1993.

[19] NuDat, NuDat2 database (2.6), in, National Nuclear Data Center, Brookhaven National Laboratory, 2014.

[20] B. Pritychenko, A. Sonzogni, Q-value calculator, in, NNDC, Brookhaven National Laboratory, 2003.

[21] A.J. Koning, S. Hilaire, M.C. Duijvestijn, TALYS-1.0, in: O. Bersillon, F. Gunsing, E. Bauge, R. Jacqmin, S. Leray (Eds.) International Conference on Nuclear Data for Science and Technology, EDP Sciences, Nice, France, 2007, pp. 211.

[22] A.J. Koning, D. Rochman, J.C. Sublet, TENDL-2017 TALYS-based evaluated nuclear data library, in, 2017.

[23] M. Herman, R. Capote, B.V. Carlson, P. Oblozinsky, M. Sin, A. Trkov, H. Wienke, V. Zerkin, EMPIRE: Nuclear reaction model code system for data evaluation, Nucl. Data Sheets, 108 (2007) 2655-2715.

[24] M. Herman, R. Capote, M. Sin, A. Trkov, B. Carlson, P. Oblozinsky, C. Mattoon, H. Wienke, S. Hoblit, Y.-S. Cho, V. Plujko, V. Zerkin, Nuclear Reaction Model Code EMPIRE-3.2 (Malta), in, http://www.nndc.bnl.gov/empire/index.html, 2012.

[25] R. Capote, M. Herman, P. Oblozinsky, P.G. Young, S. Goriely, T. Belgya, A.V. Ignatyuk, A.J. Koning, S. Hilaire, V.A. Plujko, M. Avrigeanu, O. Bersillon, M.B. Chadwick, T. Fukahori, Z. Ge, Y. Han, S. Kailas, J. Kopecky, V.M. Maslov, G. Reffo, M. Sin, E.S. Soukhovitskii, P. Talou, Reference Input Parameter Library (RIPL-3), Nucl. Data Sheets, 110 (2009) 3107-3214.

[26] K. Abe, A. Iizuka, A. Hasegawa, S. Morozumi, Induced Radioactivity of Component Materials by 16-Mev Protons and 30-Mev Alpha-Particles, J Nucl Mater, 123 (1984) 972-976.

[27] F. Ditrói, P. Fehsenfeld, A.S. Khanna, I. Konstantinov, I. Majhunka, P.M. Racolta, T. Sauvage, J. Thereska, The thin layer activation method and its applications in industry, in:  IAEA TECDOC-924, Vienna, 1997.

[28] F. Ditrói, S. Takács, F. Tárkányi, E. Corniani, R.W. Smith, M. Jech, T. Wopelka, Sub-micron wear measurement using activities under the free handling limit, J. Radioanal. Nucl. Chem., 292 (2012) 1147-1152.